\documentclass[runningheads]{llncs}

\usepackage[parfill]{parskip}    
\usepackage[T1]{fontenc}
\usepackage{float}
\usepackage{graphicx}
\makeatletter
\def\subsubsection{\@startsection{subsubsection}{3}%
  \z@{.5\linespacing\@plus.7\linespacing}{.1\linespacing}%
  {\normalfont\itshape}}
\makeatother

\begin{document}
\title{Learned Local Attention Maps for Synthesising  Vessel Segmentations from T2 MRI}
\titlerunning{Learned Local Attention Maps for Synthesising  Vessel Segmentations}
%
\author{Yash Deo\inst{1} \and
Rodrigo Bonazzola\inst{1} \and
Haoran Dou\inst{1} \and
Yan Xia\inst{1} \and
Tianyou Wei\inst{1} \and
Nishant Ravikumar\inst{1,2} \and 
Alejandro F.~Frangi\inst{1,2,3,4,5} \and 
Toni Lassila\inst{1,2}
}
%
%
\institute{Centre for Computational Imaging and Simulation Technologies in Biomedicine (CISTIB), School of Computing and School of Medicine, University of Leeds, Leeds, UK 
\and
NIHR Leeds Biomedical Research Centre (BRC), Leeds, UK
\and
Alan Turing Institute, London, UK
\and
Medical Imaging Research Center (MIRC), Electrical Engineering and Cardiovascular Sciences Departments, KU Leuven, Leuven, Belgium
\and
Division of Informatics, Imaging and Data Science, Schools of Computer Science
and Health Sciences, University of Manchester, Manchester, UK
}
\maketitle              
\begin{abstract}
Magnetic resonance angiography (MRA) is an imaging modality for visualising blood vessels. It is useful for several diagnostic applications and for assessing the risk of adverse events such as haemorrhagic stroke (resulting from the rupture of aneurysms in blood vessels). However, MRAs are not acquired routinely, hence, an approach to synthesise blood vessel segmentations from  more routinely acquired MR contrasts such as T1 and T2, would be useful. We present an encoder-decoder model for synthesising segmentations of the main cerebral arteries in the circle of Willis (CoW) from only T2 MRI. We propose a two-phase multi-objective learning approach, which captures both global and local features. It uses learned local attention maps generated by dilating the segmentation labels, which forces the network to only extract information from the T2 MRI relevant to synthesising the CoW. Our synthetic vessel segmentations generated from only T2 MRI achieved a mean Dice score of $0.79 \pm 0.03$ in testing, compared to state-of-the-art segmentation networks such as transformer U-Net ($0.71 \pm 0.04$) and nnU-net($0.68 \pm 0.05$), while using only a fraction of the parameters. The main qualitative difference between our synthetic vessel segmentations and the comparative models was in the sharper resolution of the CoW vessel segments, especially in the posterior circulation.

\end{abstract}
\section{Introduction}

A magnetic resonance angiogram (MRA) contains vital information for visualising the brain vasculature, which includes an anastomotic ring of arteries located at the base of the brain called the circle of Willis (CoW). Multiple different topological variants of the CoW exist in the general population, and certain variants of the CoW can lead to worse outcomes following a stroke~\cite{Lin2022}. To that end, it would be useful to visualise the main cerebral blood vessels in large imaging datasets and identify them by CoW phenotype to understand their relevance to stroke in the general population. Vessel segmentation from MRA is a well-studied problem with state-of-the-art methods achieving high quality vessel segmentation results~\cite{LIN2023107355} with Dice scores as high as 0.91~\cite{Xiao2022}. However, as MRA acquisition may require the injection of contrast agents and has longer acquisition times, it is not commonly available in population imaging studies. T1- and T2-weighted MRI scans are the most common MR imaging modalities available and are used to study the presence of lesions or other abnormal structures in the brain. While the blood vessels are not explicitly visible in these modalities, they contain latent information that can be used to synthesise the major vessels in the brain.

Generative adversarial neural networks~\cite{GAN} (GANNs) have seen remarkable success in the field of image synthesis, with networks like pix2pix~\cite{pix2pix} achieving impressive results in paired image-to-image synthesis. GANNs have also been widely used in medical image synthesis in various use cases such as generating T1, T2, and FLAIR images of the brain using Wasserstein-GANNs~\cite{Wassgan}. Progressively growing GANNs~\cite{Beers} have been used for the generation of retinal fundus and brain images. Previous works on brain MRA synthesis used SGAN~\cite{SGAN} to generate MRA from paired T1 and T2 images, or used starGAN~\cite{star} to synthesise MRA given T1, T2 and/or a PD-weighted MRI as input. GANN-based approaches such as vox2vox~\cite{vox2vox} have been used to synthesise segmentations of brain tumour directly from T1, T2, Gadolinium-enhanced T1, and T2 FLAIR modalities. Most GANN based approaches synthesise MRA from multiple other MR modalities, and then require the use of a separate segmentation algorithm, such as U-net (which is popularly accepted as baseline), to segment the brain vascular structures from the synthesised MRA. As the brain vessels form a very small portion of the MRA image, attention mechanisms were introduced to the segmentation algorithms to more accurately capture the small vessels. This has been achieved in networks such as Attention U-Net~\cite{aunet} or more recently transformer based networks such as TransU-Net ~\cite{tunet}.

In spite of their successes, GANs and transformers are complex models with tens or hundreds of millions of parameters that can be notoriously hard to train. On top of that, GANNs tend to produce phantoms (non-existent image features), especially when dealing with very high-resolution images with intrinsic detail arising from medical imaging~\cite{gandr}. To alleviate these issues, we propose multi-task  learnable localised attention maps to directly generate vessel segmentations based on a U-Net architecture, which can capture both global and local features from the input domain. Our method requires only the T2 modality as in input, which eliminates the need of multiple input modalities. The learned local attention maps enable the trained model to only look for vessels in specific parts of the image, which drastically decreases the number of parameters required to train the synthesis network. Our model consequently synthesises more accurate CoW segmentations with fewer parameters than competing GANN-based approaches.
 
\section{Methodology}

We propose a deep convolutional encoder-decoder model, which is trained with two-phase multi-task learning. At training time, paired T2 images and ground-truth MRA segmentations are available. Our encoder-decoder network captures both global information (by encoding input images into a latent space) and local information (by learning soft attention maps for brain vessels based on MRA segmentations) from the given input images. We train the model using multi-task learning in two phases, where a learned local attention map learns where on the T2 image the vessels are most likely located to improve the synthesised vessel segmentation masks. At run-time, the model efficiently synthesises brain vessel segmentation masks from only T2 images. 

\subsection{Data and Pre-processing}

\begin{figure}[ht!]
\includegraphics[width=\textwidth]{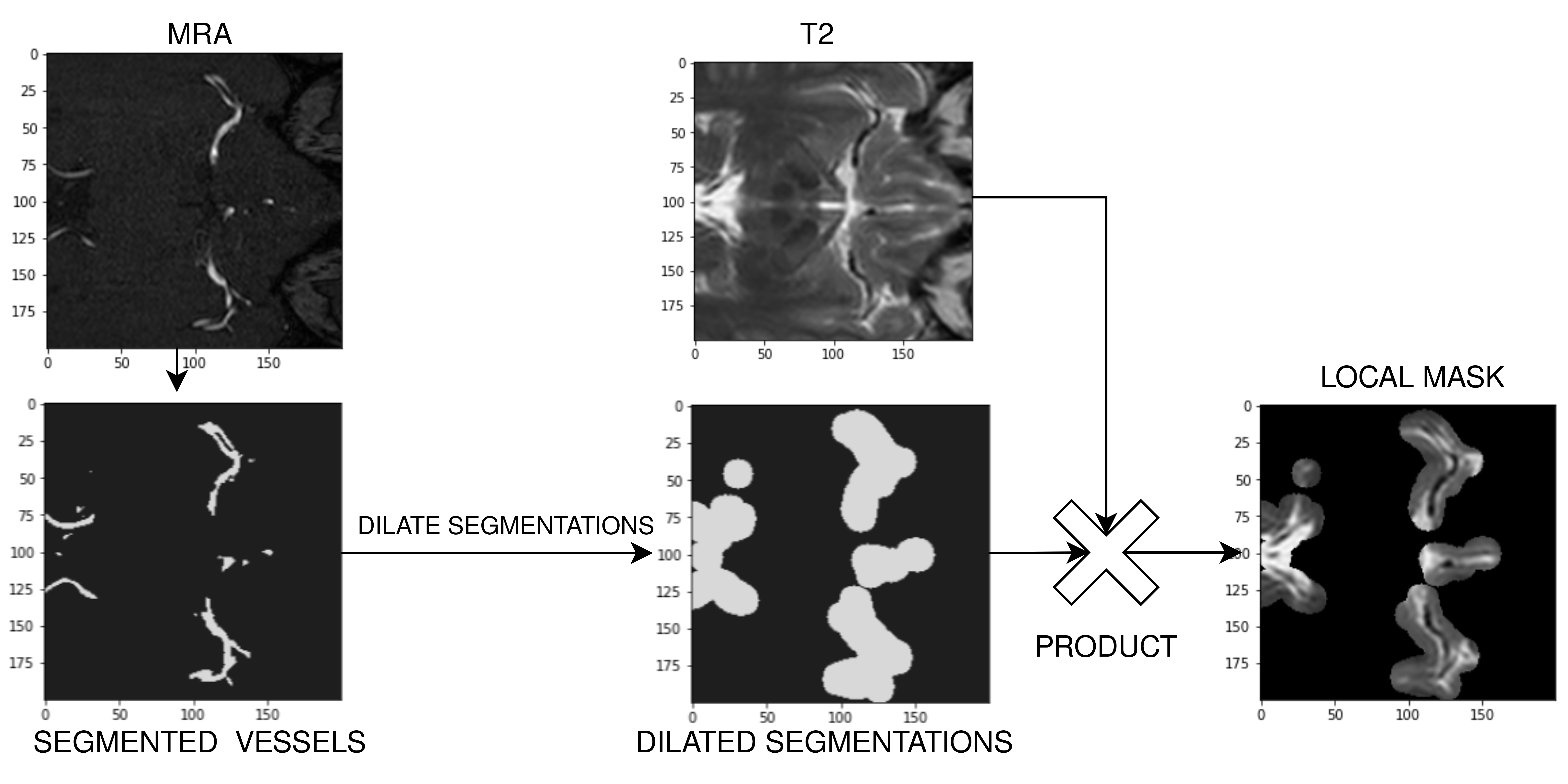}
\caption{Process for the generation of the local attention masks. Vessel segmentations are generated from the MRA and dilated. We then multiply this dilation with the corresponding T2 slice to create the mask. } 
\label{fig:1}
\end{figure}

The model was trained on the IXI dataset~\cite{IXI} using the 3T scans acquired at Hammersmith Hospital, and includes paired T2 and MRA scans of 181 patients. The T2 and MRA images were first registered using rigid registration. The images were centered, cropped from $512\times512$ to $400\times400$, and intensity-normalised. Ground-truth segmentations were then generated from the MRA images for each corresponding T2 slice using a residual U-Net~\cite{Kerfoot18}. The segmentations were then dilated to form a binary mask and multiplied pixelwise with the corresponding T2 slice to create the ground truth local attention map (see Fig.~\ref{fig:1})

\subsection{Network Architecture}

 The proposed model follows the general architecture of the pix2pix-model~\cite{pix2pix} with one encoder branch and two output branches (Fig.~\ref{fig:2}). The encoder branch combines U-net and Resnet~\cite{resnet} architectures with a latent space consisting of three consecutive residual blocks, similar to the vox2vox-model~\cite{vox2vox}. The encoder has four convolution + max-pooling -blocks, where each block consists of three strided convolution layers followed by a max-pooling layer. Each convolution layer is followed by an instance-normalisation -layer. The latent space branches out into two output branches: the decoding branch and the synthesis branch. In case of multiple input modalities (eg. T1 + T2) we have a separate decoding branch for each modality. The output branches have the same structure as the encoding branch with the max-pooling layers replaced by up-sampling layers and with skip connections from corresponding encoding blocks. The first convolution block of the synthesis branch receives a skip connection from both the corresponding encoder branch and the decoder branch.
 
\paragraph{Local Attention Mask} The output of the segmentation branch consists of fine vessel information. The small dimensions of the vessels make the segmentation masks unsuitable for generating the local attention maps. For this reason, we dilate these vessel segments to 10 pixels in each direction to create a local attention mask. The optimal dilation width was found through experimentation as shown in Table \ref{table:3}. We then perform pixel-wise multiplication of this local attention mask with the output of the decoder to generate a local attention map as shown in Fig.~\ref{fig:1}. This local attention map is compared to the ground truth local attention maps during model training to calculate loss. This dependency between these two tasks adds a collaborative element between what would otherwise be two contrastive tasks. The use of a local attention mask forces the network to learn from a very small portion of the input image, which contains information about the blood vessels and ignore the rest of the image. This property allows us to greatly reduce the number of parameters required to train the model.


\subsection{Training and Losses}

The network is trained in two phases to effectively capture both the global and local features required to synthesise the vessels from T2 images.\\ 

\begin{figure}[h!]

\includegraphics[width=\textwidth]{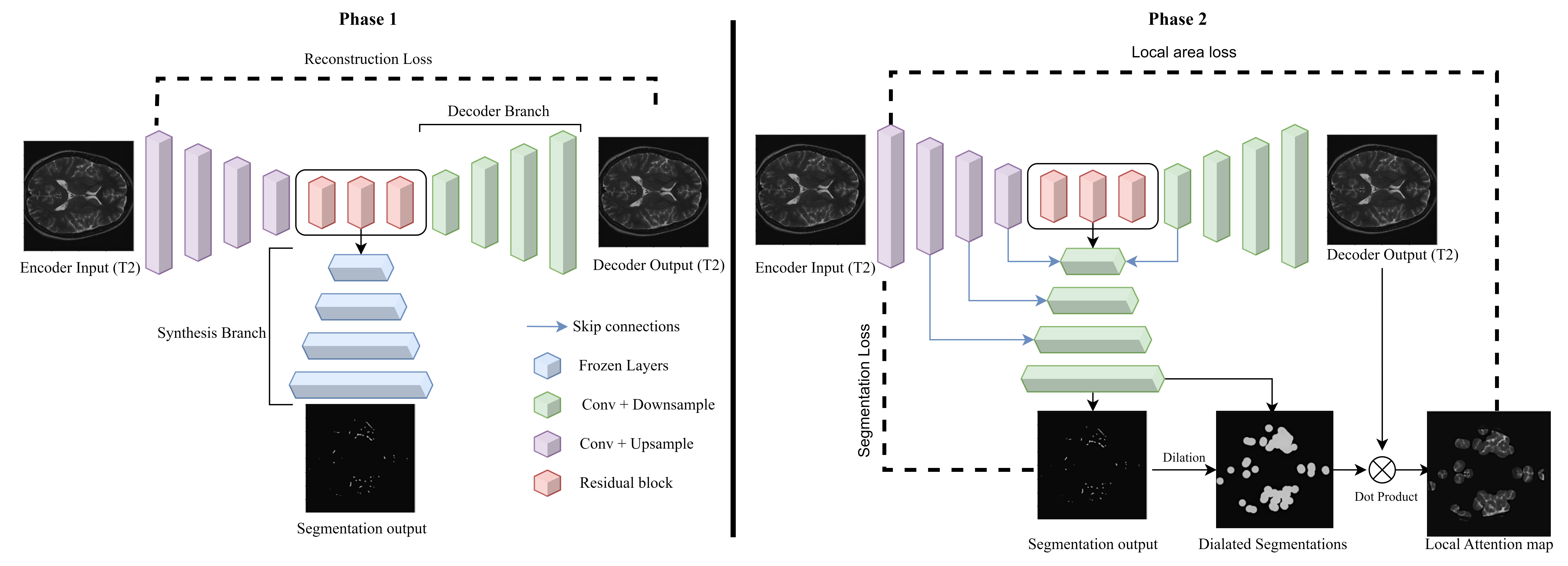}
\caption{Overview of our network architecture. The encoder takes T2-weighted MRI as input and compresses it into a latent space. The latent space branches out into the decoding branch, which reconstructs the input, and the synthesis branch, which generates the segmentation.} \label{fig:2}
\end{figure}

\paragraph{Phase 1:}

We pre-train the network on T2 images by freezing the synthesis branch and only training the decoder branch, effectively training an autoencoder for T2 images. The network is trained with an early stopping criteria based on the loss slope. The only loss calculated in this stage is the T2 reconstruction loss from the decoder branch.The loss function used is L1 and is specified below where $X_{T_{2}}$ is the ground truth T2 image and $\hat{X}_{T_2}$ is the generated T2 image:

\begin{equation}
\mathcal{L}_{\textrm{phase}\,1} = \textrm{MAE}(X_{T_{2}},\hat{X}_{T_2})
\end{equation}

\paragraph{Phase 2:}

After we finish the pre-training step, we unfreeze the synthesis branch and train it in conjunction with the decoder branch. Although the decoder branch is being trained in this step, the loss calculated for this branch is not the reconstruction loss but local loss, which is calculated over the dot product of the output of the decoder branch and the dilated segmentation obtained from the output of the synthesis branch.

In order to train these two contrasting branches together, we tested our model with various multi-task learning (MTL) approaches: Nash-MTL~\cite{NashMTL} (average Dice after evaluation 0.76), CAGrad~\cite{cagrad} (average Dice after evaluation 0.74), and uncertainty-based MTL ~\cite{umtl} (average Dice after evaluation 0.79). The best performing version was the uncertainty-based MTL, where both the losses are weighted based on the assumption of homoscedastic uncertainty for each task. The loss function for our multi-output model is described in (\ref{eq:loss}), where $W$ are the model parameters and we interpret minimising the loss with respect to  $\sigma_1$ and  $\sigma_2$ as learning the relative weights for the losses $\mathcal{L}_{\textrm{seg}}$ and $\mathcal{L}_{\textrm{loc}}$ adaptively. We used Dice score as the loss for $\mathcal{L}_{\textrm{seg}}$ and MAE as the loss for $\mathcal{L}_{\textrm{loc}}$ 

\begin{equation} \label{eq:loss}
\mathcal{L}_{\textrm{phase}\,2}=\frac{1}{2 \sigma_1^2} \mathcal{L}_{\textrm{seg}}(\mathbf{W})+\frac{1}{2 \sigma_2^2} \mathcal{L}_{\textrm{loc}}(\mathbf{W})+\log \sigma_1 \sigma_2
\end{equation}




\section{Experiments and results}

\subsection{Implementation Details}

All the models were implemented in TensorFlow 2.8 and Pytorch (for nnU-Net) and Python 3.  Out of the 181 cases in the dataset we used 150 for training and 31 for testing and validation. All the models were pre-trained on T2 images and grid search was used to optimise the following hyperparameters: (1) batch size, (2) learning rate, (3) number of epochs, and (4) momentum. To train the transformer network, we first used the parameters recommended in \cite{tunet} and applied further fine-tuning of the parameters to achieve comparative performance in the segmentation task. 
\begin{figure}[ht!]

\includegraphics[width=\textwidth]{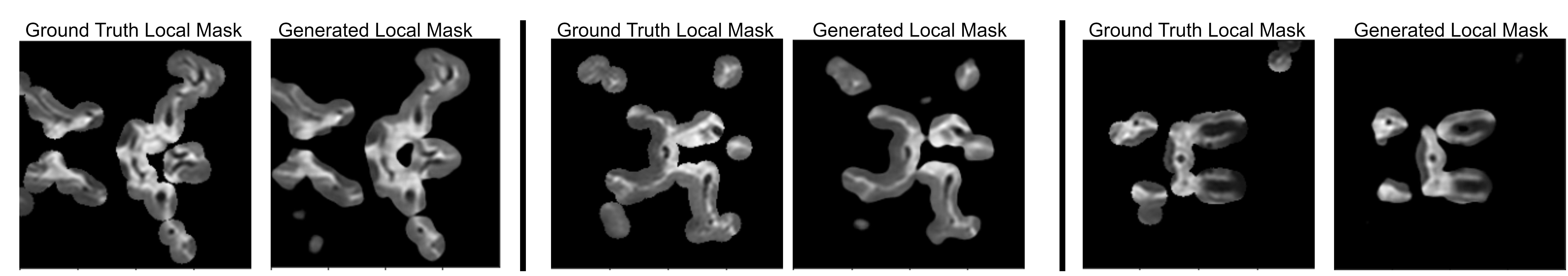}
\caption{Local attention maps learned by the network compared against the ground truth local attention maps.} \label{fig:2}
\end{figure}


To evaluate the results of our model against other methods, we used the segmentation metrics of Dice score and Hausdorff distance (hd95). The results were averaged over the 3D volumes of the 11 leave-out cases and are shown in Table \ref{table:2}. Our method clearly outperforms conventional GANN-based synthesis methods, such as vox2vox, and also performs slightly better than state-of-the-art segmentation models like transformer U-Net~\cite{tunet} and nnU-net~\cite{nnunet21}, while also being easier to train with fewer trainable parameters. We experimented with training our model with different input modalities, which showed that using only T1 as an input had the worst performance (average dice 0.64 $\pm 0.04$) while the performance of using only T2 (average dice 0.79 $\pm 0.04$) and both T1 + T2 (average dice 0.78 $\pm 0.05$) was essentially the same, with T1 + T2 requiring additional parameters (33.4 million) compared to using just T2 (26.7 million) as we would need an additional decoding branch for the T1 decoder. A crucial hyperparameter in our model is the dilation width of the segmentations to generate the local attention maps, which was optimised in a separate experiment. (Table \ref{table:3}).  

\begin{table}[]
\centering
\caption{Difference in loss with different values of dilation for the local attention mask}
\label{table:3}
\begin{tabular}{l|c|c}
Attention mechanism used       &  Dice (95\% CI)  & Area covered by mask\\
\hline
No local attention mask       & 0.62 $\pm 0.04$    & NA \\ 
Mask with no dilation         & 0.59 $\pm 0.04$ & 1.5\% \\
Mask with dilation by 5 pixels      & 0.74 $\pm 0.03$ & 8.5\%\\
Mask with dilation by 10 pixels     & 0.79 $\pm 0.03$ & 18\%\\
Mask with dilation by 15 pixels      & 0.75 $\pm 0.02$ &  28\%\\
Mask with dilation by 20 pixels     & 0.75 $\pm 0.03$ & 37\%\\
\end{tabular}
\end{table}

\begin{table}[ht!]
\centering
\caption{Accuracy of synthesised vessel segmentation masks in a test set of $11$ leave-out cases}
\label{table:2}
\begin{tabular}{l|c|c|c|l}
Model        & Model                    & Dice       & HD95       &  Model Type\\
             &  params. ($\times 10^6$) &  (95\% CI) &  (95\% CI) &   \\
\hline
Our model         & $26.7$  & 0.79 $\pm 0.03$  & 9.1  $\pm 0.5$  & Segmentation/synthesis\\
Transformer U-Net~\cite{tunet}  & $105.8$  & 0.71 $\pm 0.04$  & 10.4 $\pm 0.5$   & Segmentation    \\
nnU-Net~\cite{nnunet21}  & $127.8$  & 0.68 $\pm 0.03$  & 9.3 $\pm 0.4$   & Segmentation    \\
Vox2vox~\cite{vox2vox}     & $78.8$ & 0.67 $\pm 0.05$  & 17.2 $\pm 1.4$   & Segmentation/synthesis  \\
Pix2pix~\cite{pix2pix}     & $36.9$   & 0.55 $\pm 0.04$  & 23.1 $\pm 3.0$  & Synthesis\\
U-Net~\cite{unet}  (base) & $9.1$  & 0.57 $\pm 0.05$   & 42.6 $\pm 4.2$  & Segmentation       \\
\end{tabular}
\end{table}

\subsection{Qualitative Results}

\begin{figure}[t!]
\includegraphics[width=\textwidth]{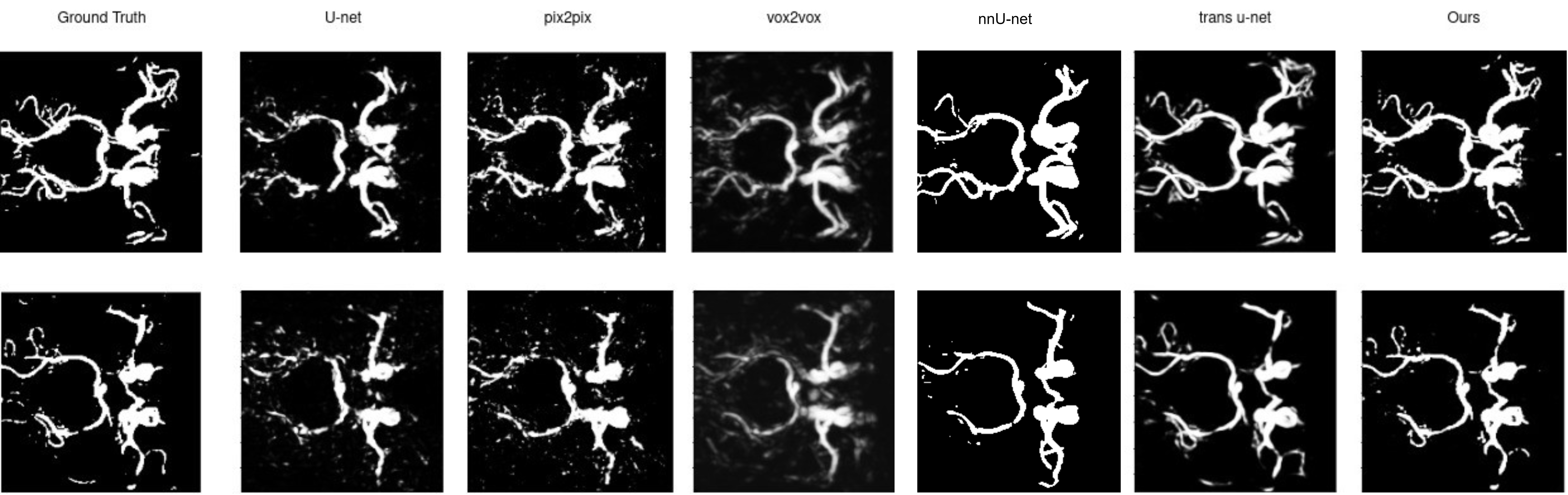}
\caption{CoW synthesis results compared between models. Pix2pix and U-Net are able to capture the overall structure of the Cow but with a lot of noise. Vox2vox performs comparatively better, but still suffers from noise in the outputs. NnU-Net, Transformer U-Net and our method show good results with our method capturing more details and dealing better with noise.} 

\label{fig:3}
\includegraphics[width=\textwidth]{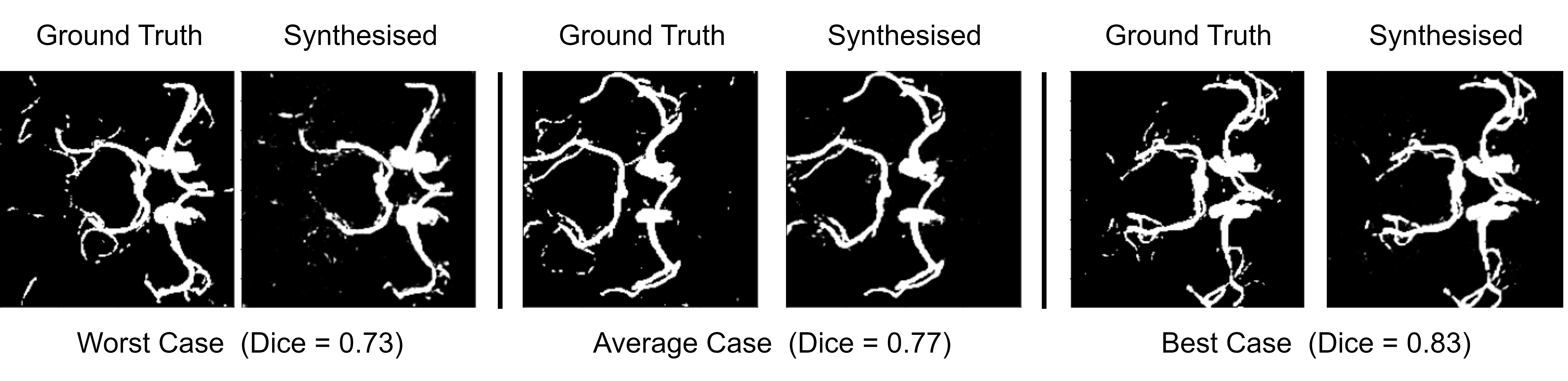}
\caption{CoW synthesis results for the average case, the best case, and the worst case in our unseen test set. } 
\label{fig:4}
\end{figure}

Fig.~\ref{fig:3} shows a qualitative comparison of our method against pix2pix, vox2vox, U-Net, nnU-net, and transformer U-Net for two samples from the unseen test set. It can be observed that pix2pix and the base U-Net are only able to capture the overall structure of the CoW with a lot of noise. The vox2vox model synthesises the vessels slightly better, but is still unable to capture the finer details and suffers from noise. The nnU-net and  transformer U-Net are able to synthesise the vessels with high quality, but struggle to synthesise smaller vessels such as the posterior communicating arteries (PComA) in the first case. An interesting observation can be made in the second case, where the ground truth has faults in the segmentation (especially in the posterior circulation). The Transformer U-Net, nnU-net, and our model attempt to fix these faults by synthesising a continuous PCA, but our model does better in restoring vessel continuity. Fig.~\ref{fig:4} shows the CoW synthesis results for the best case, worst case, and median case scenarios. It can be observed that in the worst case the model struggles to synthesise the smaller vessels towards the end of the posterior cerebral circulation, whereas in the median case scenario most of the major vessels are synthesised with only the small PComA artery missing. The best case is that all the major arteries of the CoW are synthesised while also removing noise from the input image.



\subsection{Limitations}

While our method outperforms state-of-the-art approaches with a much smaller number of trainable parameters and is able to generated the complete structure of the CoW, it can be seen that in come cases the model can struggle to generate some of the finer vessels branching from the main arteries (especially the posterior communicating arteries). This could be either because the input data is of insufficient resolution (T2 images were acquired at 3T) or because the T2 modality does not contain information that could be used to synthesise the anterior circulation. It is possible that additional MR modalities, such as multi-view T1, or a fully-3D neural network architecture could add more information about the posterior and anterior vessels and recover a complete CoW.



\section{Conclusion}

We proposed a multi-output encoder-decoder -based network that learned to effectively synthesise vessels from only T2-weighted MRI using local attention maps and multi-task learning. The qualitative and quantitative results show that our method outperformed both the state-of-the-art and conventional segmentation/synthesis algorithms, while at the same time being easier to train with fewer parameters. In future work, we are extending our model to a fully 3D synthesis model to achieve better connectivity of the CoW structure.

\section*{Acknowledgement}

This research was partially supported by the National Institute for Health and Care Research (NIHR) Leeds Biomedical Research Centre (BRC) and the Royal Academy of Engineering Chair in Emerging Technologies (CiET1919/19).

%

\newpage

\bibliographystyle{splncs04}
\bibliography{mybib}

\begin{thebibliography}{10}
\providecommand{\url}[1]{\texttt{#1}}
\providecommand{\urlprefix}{URL }
\providecommand{\doi}[1]{https://doi.org/#1}

\bibitem{Beers}
Beers, A., Brown, J., Chang, K., Campbell, J.P., Ostmo, S., Chiang, M.F.,
  Kalpathy-Cramer, J.: High-resolution medical image synthesis using
  progressively grown generative adversarial networks. arXiv preprint
  arXiv:1805.03144  (2018)

\bibitem{tunet}
Chen, J., Lu, Y., Yu, Q., Luo, X., Adeli, E., Wang, Y., Lu, L., Yuille, A.L.,
  Zhou, Y.: Transunet: {T}ransformers make strong encoders for medical image
  segmentation. arXiv preprint arXiv:2102.04306  (2021)

\bibitem{vox2vox}
Cirillo, M.D., Abramian, D., Eklund, A.: Vox2{V}ox: {3D-GAN} for brain tumour
  segmentation. In: Brainlesion: {G}lioma, {M}ultiple {S}clerosis, {S}troke and
  {T}raumatic {B}rain {I}njuries: 6th {I}nternational {W}orkshop, {B}rain{L}es
  2020, Held in Conjunction with {MICCAI} 2020, {L}ima, {P}eru, {O}ctober 4,
  2020, Revised Selected Papers, {Part} I.6. pp. 274--284. Springer (2021)

\bibitem{GAN}
Goodfellow, I., Pouget-Abadie, J., Mirza, M., Xu, B., Warde-Farley, D., Ozair,
  S., Courville, A., Bengio, Y.: Generative adversarial networks. Commun. ACM
  \textbf{63(11)},  139--144 (2020)

\bibitem{Wassgan}
Han, C., Hayashi, H., Rundo, L., Araki, R., Shimoda, W., Muramatsu, S.,
  Furukawa, Y., Mauri, G., Nakayama, H.: {GAN}-based synthetic brain {MR} image
  generation. In: 2018 IEEE 15th Intl. Sympos. Biomed. Imag. (ISBI 2018). pp.
  734--738. IEEE (2018)

\bibitem{resnet}
He, K., Zhang, X., Ren, S., Sun, J.: Deep residual learning for image
  recognition. In: Proc. IEEE Conf. Comput. Vis. Pattern Recog. pp. 770--778
  (2016)

\bibitem{IXI}
{Information eXtraction from Images Consortium}: {IXI} dataset -- brain
  development. \url{https://brain-development.org/ixi-dataset/}, accessed:
  2023-02-14

\bibitem{nnunet21}
Isensee, F., Jaeger, P.F., Kohl, S.A., Petersen, J., Maier-Hein, K.H.:
  nn{U}-{N}et: a self-configuring method for deep learning-based biomedical
  image segmentation. Nat. Methods  \textbf{18}(2),  203--211 (2021)

\bibitem{pix2pix}
Isola, P., Zhu, J.Y., Zhou, T., Efros, A.A.: Image-to-image translation with
  conditional adversarial networks. In: Proc. IEEE Conf. Comput. Vis. Pattern
  Recog. pp. 1125--1134 (2017)

\bibitem{umtl}
Kendall, A., Gal, Y., Cipolla, R.: Multi-task learning using uncertainty to
  weigh losses for scene geometry and semantics. In: Proc. IEEE Conf. Comput.
  Vis. Pattern Recog. pp. 7482--7491 (2018)

\bibitem{Kerfoot18}
Kerfoot, E., Clough, J., Oksuz, I., Lee, J., King, A.P., Schnabel, J.A.:
  Left-ventricle quantification using residual {U-Net}. In: Statistical Atlases
  and Computational Models of the Heart. Atrial Segmentation and LV
  Quantification Challenges: 9th International Workshop, STACOM 2018, Held in
  Conjunction with MICCAI 2018, Granada, Spain, September 16, 2018, Revised
  Selected Papers 9. pp. 371--380. Springer (2019)

\bibitem{Lin2022}
Lin, E., Kamel, H., Gupta, A., RoyChoudhury, A., Girgis, P., Glodzik, L.:
  Incomplete circle of {W}illis variants and stroke outcome. Eur. J. Radiol.
  \textbf{153},  110383 (2022)

\bibitem{LIN2023107355}
Lin, F., Xia, Y., Song, S., Ravikumar, N., Frangi, A.F.: High-throughput 3dra
  segmentation of brain vasculature and aneurysms using deep learning. Computer
  Methods and Programs in Biomedicine  \textbf{230},  107355 (2023).
  \doi{https://doi.org/10.1016/j.cmpb.2023.107355},
  \url{https://www.sciencedirect.com/science/article/pii/S0169260723000226}

\bibitem{cagrad}
Liu, B., Liu, X., Jin, X., Stone, P., Liu, Q.: Conflict-averse gradient descent
  for multi-task learning. Adv. Neural Inf. Process. Syst.  \textbf{34},
  18878--18890 (2021)

\bibitem{NashMTL}
Navon, A., Shamsian, A., Achituve, I., Maron, H., Kawaguchi, K., Chechik, G.,
  Fetaya, E.: Multi-task learning as a bargaining game. arXiv preprint
  arXiv:2202.01017  (2022)

\bibitem{aunet}
Oktay, O., Schlemper, J., Folgoc, L.L., Lee, M., Heinrich, M., Misawa, K.,
  Mori, K., McDonagh, S., Hammerla, N.Y., Kainz, B., et~al.: Attention {U-net}:
  Learning where to look for the pancreas. arXiv preprint arXiv:1804.03999
  (2018)

\bibitem{SGAN}
Olut, S., Sahin, Y.H., Demir, U., Unal, G.: Generative adversarial training for
  {MRA} image synthesis using multi-contrast {MRI}. In: {PR}edictive
  {I}ntelligence in {ME}dicine: {F}irst {I}nternational {W}orkshop, {PRIME}
  2018, Held in Conjunction with {MICCAI} 2018, {G}ranada, {S}pain, {S}eptember
  16, 2018, {P}roceedings Vol.1. pp. 147--154. Springer (2018)

\bibitem{unet}
Ronneberger, O., Fischer, P., Brox, T.: {U-net}: {C}onvolutional networks for
  biomedical image segmentation. In: Medical Image Computing and
  Computer-Assisted Intervention--MICCAI 2015: 18th International Conference,
  Munich, Germany, October 5-9, 2015, Proceedings, Part III 18. pp. 234--241.
  Springer (2015)

\bibitem{star}
Sohail, M., Riaz, M.N., Wu, J., Long, C., Li, S.: Unpaired multi-contrast {MR}
  image synthesis using generative adversarial networks. In: Simulation and
  {S}ynthesis in {M}edical {I}maging: 4th {I}nternational {W}orkshop, {SASHIMI}
  2019, Held in Conjunction with {MICCAI} 2019, {S}henzhen, {C}hina, {O}ctober
  13, 2019, {P}roceedings. pp. 22--31. Springer (2019)

\bibitem{Xiao2022}
Xiao, R., Chen, C., Zou, H., Luo, Y., Wang, J., Zha, M., Yu, M.: Segmentation
  of cerebrovascular anatomy from {TOF-MRA} using length-strained enhancement
  and random walker. Biomed Res. Int.  \textbf{2020},  9347215 (2020)

\bibitem{gandr}
Yu, B., Wang, Y., Wang, L., Shen, D., Zhou, L.: Medical image synthesis via
  deep learning. In: Deep Learning in Medical Image Analysis: Challenges and
  Applications, Advances in Experimental Medicine and Biology, vol.~1213, pp.
  23--44. Springer (2020)

\end{thebibliography}
\end{document}